\documentclass[unnumsec,webpdf,modern,large]{oup-authoring-template}%

\graphicspath{{Fig/}}

\theoremstyle{thmstyleone}%
\theoremstyle{thmstyletwo}%
\theoremstyle{thmstylethree}%

\usepackage{float}

\usepackage{xurl} 

\begin{document}

\journaltitle{}
\DOI{}
\copyrightyear{}
\pubyear{}
\vol{}
\issue{}
\access{}
\appnotes{}

\firstpage{1}


\title[Dartmouth Flitzing Culture and Artificial Intelligence]{Human-AI Interaction Mediated by Courtship Language Culture: Computational Statistics and Probabilistic Modeling}

\author[1,2$\dagger$$\ast$]{Linh Huynh\ORCID{0000-0003-0540-6811}}
\author[1$\dagger$]{Nick Cyriax}
\author[1a]{Jackson George}
\author[1a]{Stephanie Zoltick}
\author[1]{Zachariah Yusaf}

\address[1]{\orgdiv{Department of Mathematics}, \orgname{Dartmouth College}, \orgaddress{\street{Hanover}, \state{NH} \postcode{03755}, \country{USA}}}
\address[2]{\orgdiv{Honors College Applied Mathematics}, \orgname{Texas Tech University}, \orgaddress{\street{Lubbock}, \state{TX} \postcode{79409}, \country{USA}}}
\address[$\dagger$]{L.H. and N.C. have contributed equally to this work as co-first authors.}
\address[a]{J.G. and S.Z. contributed equally to this work as co-second authors.}
\corresp[$\ast$]{To whom correspondence should be addressed: \href{mailto:linh.n.huynh@ttu.edu}{linh.n.huynh@ttu.edu}}



\abstract{Large language models are entering living cultural practices, including intimate ones. We study a collegiate courtship tradition at Dartmouth College called flitzing, the tradition of sending rhymed flirtatious emails to ask peers out, as a measurable microcosm of a culture absorbing artificial intelligence. We combine four sources of evidence: human evaluation surveys in two waves (2025 and 2026); information retrieval over 33 human-written and 5,610 AI-generated flitzes spanning 17 sampling temperatures; evaluation of the same survey by three generative language models; and simulation of a coupled multi-species bipartite spin glass model of people and texts. Four main findings emerge. First, sampling temperature induces a phase transition in diversity: below the transition, AI-generated flitzes are more homogeneous than the human, while above it they decohere, and human baselines of diversity remain stable across years, indicating persistent collective structure in human creativity. Second, human judges conflate perceived quality with human authorship, whereas AI judges detect provenance almost perfectly yet still rate human-written flitzes higher. Third, the transition window, with temperature between 1.25 and 1.38, is consistent across biased survey data, unbiased text statistics, and the mathematical model, which also reproduces the observed clustering. Fourth, detection ability generalizes beyond the local culture, with Dartmouth and non-Dartmouth respondents performing comparably, and strengthens over time, with 2026 respondents above chance at every temperature. We conclude that AI participation in a tight cultural genre is regime dependent: conformist and homogenizing below criticality, detectable and disruptive above it.}



\otherabstract[Significance statement]{Courtship writing is among the most personal of cultural practices, and generative artificial intelligence is increasingly involved in such practices. Using a codified campus tradition, Dartmouth flitzing, we measure how a generative language model behaves inside a tight living culture as its sampling temperature varies. We find a sharp transition separating a conformist regime, in which AI output passes as human while homogenizing the genre, from a decohered regime, in which it is easily detected. Human judges mistake quality for humanity; AI judges do so significantly less, yet they prefer human writing. The framework, combining surveys, text statistics, and spin glass theory, offers a quantitative approach for studying cultural evolution under artificial intelligence.}

\maketitle

\section{Introduction}
When artificial intelligence becomes embedded in a human social network, how does it alter the collective dynamics of cultural evolution and drive transitions between distinct behavioral regimes? In this paper, we address this question quantitatively in a living courtship tradition at Dartmouth College, where students send one another ``flitzes,'' which are rhymed flirtatious invitations composed in a recognizable house style and circulated over the campus email system \cite{thedartmouth2022flitzing}. The practice is widespread rather than marginal. In a March 2025 campus poll of 309 students, 55\% of respondents reported having sent or received at least one flitz, and roughly a third had sent one to someone they were not dating \cite{thedartmouth2025dating}. The flitz belongs to a long tradition in which written courtship follows recognizable conventions, with the text representing the writer’s inner self \cite{lystra1989searching}. Courtship more broadly also follows learned, rule-governed scripts \cite{bailey1988front}, and campus intimacy remains shaped by such scripts even when it appears informal \cite{bogle2008hooking,wade2017american}. Courtship writing was already a strategic, mediated performance before artificial intelligence (AI) arrived, from optimized online dating profiles \cite{ellison2006managing} to the ``media ideologies'' by which students read meaning into the choice of channel itself \cite{gershon2010breakup}; meeting online is now the dominant pathway for American couples \cite{rosenfeld2019disintermediating}. AI extends this outsourcing to authorship itself: daters use large language models (LLMs) to draft self-presentations \cite{barkallah2025transparent}, perceived algorithmic authorship reduces trust and attraction \cite{wu2020online}, and at the far end of the arc people form romantic bonds with chatbots directly \cite{skjuve2021chatbot}. Because flitzing is a well-defined and still-active genre with clear conventions, it provides a useful setting for studying what happens when a generative model enters a local human tradition.

AI raises the quality of individual creativity while narrowing the collective diversity of what a population produces \cite{doshi2024generative}; co-writing with instruction-tuned models reduces content diversity across authors \cite{padmakumar2024does}, and users of LLM-assisted ideation converge without noticing that they do \cite{anderson2024homogenization}. At the level of institutions, universal adoption of a single algorithm can reduce aggregate welfare even when the algorithm outperforms each individual it replaces \cite{kleinberg2021algorithmic}. These homogenization concerns are no longer hypothetical for written genres: an excess of characteristic LLM vocabulary is measurable across millions of biomedical abstracts \cite{kobak2025delving}, and ChatGPT-preferred words are rising in human speech itself \cite{yakura2024empirical,saadat2024impact}. In this context, sampling temperature is often viewed as the LLM’s ``creativity dial,'' but empirical evidence for this belief is limited within the conventional operating range \cite{peeperkorn2024temperature,renze_guven}. Whether temperature controls a genuine tradeoff between conformity and diversity inside a real cultural practice has, to our knowledge, not been measured directly.

LLMs are trained disproportionately on text from Western, educated, industrialized, rich, and democratic (WEIRD) populations \cite{henrich2010weirdest, longpre2025bridging}; their survey responses resemble WEIRD respondents most closely, with fit degrading as cultural distance grows \cite{atari2023which}, their expressed values skew toward English-speaking and Protestant European cultural zones \cite{tao2024cultural}, and they penalize nonstandard language varieties beneath an overtly positive surface \cite{hofmann2024ai}. 
Community-driven benchmarks show that models underperform on culturally grounded scenarios \cite{shi2024culturebank}. Cultures also differ systematically in the strength of their norms and their tolerance of deviance \cite{gelfand2011differences}, and intelligent machines now act within them as generators, mediators, and selective forces of cultural evolution \cite{brinkmann2023machine}. The flitz is a tight micro-genre inside a hyper-WEIRD population, which makes it a conservative test case: the tradition sits close to the center of a frontier model's training distribution, so whatever diversity cost we measure here plausibly bounds from below the cost the same model imposes on more distant cultural practices. 

AI-mediated communication, in which an agent modifies, augments, or generates messages on a communicator’s behalf, provides the framework for the flitz experiment \cite{hancock2020ai}. Untrained evaluators struggle to distinguish AI-generated from human-generated text, performing no better than chance~\cite{clark2021all}; in some settings, AI text is even judged more human than human text \cite{jakesch2023human}. This pattern extends the Turing test tradition \cite{turing2007computing,searle1980minds} from general conversation to specific cultural settings, including poetry, reviews, and short fiction \cite{porter2024ai,kovacs2024turing,marco2024small}. Models can also reproduce group-level human response patterns when conditioned on individual personality and identity profiles \cite{argyle2023out}. Yet suspected AI origin can lower evaluations and trust \cite{dietvorst2015algorithm,wu2020online}. Here, we view human-AI interaction as a closed sociotechnical loop: models generate messages within the human tradition, people and machines evaluate them, and the resulting texts feed back into the tradition and future training data.

We mathematically model this loop with ideas from statistical mechanics, which connect individual- and group-level interactions. 
Both LLMs and the spin glass models from statistical mechanics are based on the Boltzmann distribution, also known as the softmax function of the following form:
\cite{bridle1990probabilistic,ackley1985learning}:
\begin{equation}
\mathrm{softmax}(x_i) := \frac{e^{(1/T)x_i}}{\sum_{j=1}^{N} e^{(1/T)x_j}} = \frac{e^{\beta x_i}}{\sum_{j=1}^{N} e^{\beta x_j}}, \quad \text{with } \beta := \frac{1}{T}.
\label{eq:softmax}
\end{equation}
The inverse temperature $\beta$ controls randomness and entropy in LLM output exactly as it does in a Gibbs measure over a disordered energy landscape. Treating language and social convention as ordering phenomena has an established statistical-physics tradition \cite{castellano2009statistical,loreto2011statistical}, as does treating neural networks as spin systems \cite{hopfield1982neural,amit1985spin,choromanska2015loss,bahri2020statistical}; modern attention is precisely the update rule of a continuous Hopfield network with an inverse-temperature parameter \cite{ramsauer2021hopfield}, and transformers \cite{vaswani2017attention} admit rigorous descriptions as interacting particle systems on the sphere exhibiting clustering and metastability \cite{geshkovski2025mathematical,karagodin2024clustering,chen2025quantitative,alcalde2025clustering,geshkovski2024dynamic}. Abrupt changes in LLM behavior are an active research area: claimed emergent abilities \cite{wei2022emergent} attract the critique that they are fragile under metrics that abruptly change LLM behavior \cite{schaeffer2023mirage}, decoding studies document a dichotomy between repetitive and incoherent regimes \cite{holtzman2020curious,nguyen2024turning}, and recent preprints report temperature-like phase transitions in generation directly \cite{nakaishi2024critical,arnold2024phase,aoyama2025language,sun2025phase}. Our contribution is that the setting is a living cultural practice confined to a small local location rather than benchmark language data. Moreover, we show consistency in behaviors across human data, AI data, statistical analysis, computational analysis, and mathematical model simulation.

We proceed with three complementary methodologies: (M1) statistical analysis and bias quantification of human evaluation surveys, a classic Turing test administered in 2025 and again in 2026; (M2) structural similarity analysis via term frequency-inverse document frequency (TF-IDF) cosine overlap between and within human-written and AI-generated flitz texts; and (M3) simulation of coupled multi-species bipartite Sherrington-Kirkpatrick (SK) spin glass models with Gibbs sampling. First, we find that temperature induces a phase transition in diversity: below $T \approx 1.25$ the model's flitzes are more homogeneous than the human tradition, above it they decohere, while the human baseline itself is stable across years. Second, human judges conflate perceived quality with human authorship, whereas AI judges detect provenance almost perfectly yet still rate human-written flitzes higher. Third, the transition window $T \in [1.25, 1.38]$ is consistent across biased survey data, unbiased text statistics, and the mathematical model. Fourth, the clustering structure of the data is reproduced by the model. Fifth, detection is robust across institutional space (Dartmouth versus non-Dartmouth respondents) and drifts upward across time, with 2026 respondents above chance at every temperature.

\section{Results}

\subsection{AI-performed conformity is less fragile under noise, while human creativity exhibits stable collective structure.}\label{sec:r1}

\begin{figure}[H]%
\centering
\includegraphics[width=\columnwidth]{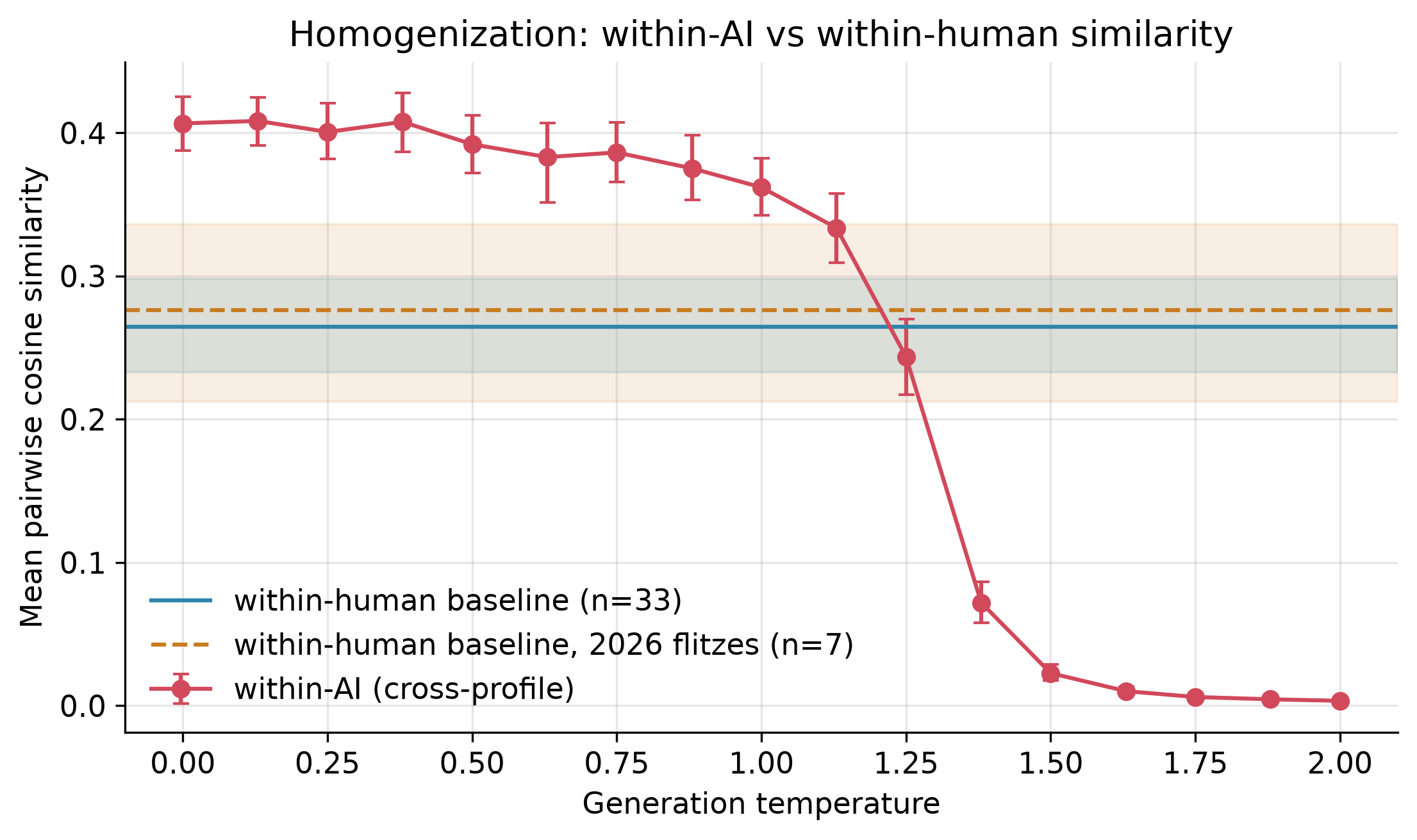}
\caption{Homogenization and diversity: within-AI versus within-human similarity. Mean pairwise TF-IDF cosine similarity between AI-generated flitzes from different virtual student profiles (red curve, error bars denote $\pm 1$ standard error) as a function of generation temperature, compared with the within-human baseline computed from the 33 human-written flitzes collected in 2025 (solid line, shaded band $\pm 1$ standard deviation) and from the 7 human-written flitzes collected in 2026 (dashed line, shaded band). Below the transition temperature the model is more homogeneous than the human tradition it imitates; above $T \approx 1.25$ the within-AI similarity collapses toward zero.}\label{fig:homog}
\end{figure}

We first ask how the internal diversity of AI-generated courtship language compares with the internal diversity of the human tradition, and how both depend on the generation temperature. For each of 17 temperature values $T \in \{0, 0.13, \ldots, 1.88, 2\}$ we compute the mean pairwise TF-IDF cosine similarity between AI-generated flitzes produced from different virtual student profiles, and we compare it against the mean pairwise similarity among the 33 human-written flitzes collected in 2025 and, separately, among the 7 human-written flitzes collected in 2026. Fig.~\ref{fig:homog} reports the result.

We make two observations. 
Firstly, There is a temperature-induced phase transition in the diversity of AI courtship language. For $T < 1.25$, within-AI similarity is approximately $0.40$, well above the human-human baseline of approximately $0.26$: the model's flitzes resemble one another substantially more than human flitzes resemble one another, even though each is conditioned on a distinct student profile. The curve declines slowly across the low-temperature regime, crosses the human band between $T = 1.13$ and $T = 1.25$, and then collapses, reaching $0.07$ at $T = 1.38$ and values below $0.03$ for $T \geq 1.5$. Because higher temperature is the standard operational proxy for creativity in LLM generation \cite{peeperkorn2024temperature,renze_guven}, this result states that when the model is made more creative it also becomes more diverse, and that the two regimes are separated by a sharp transition rather than a smooth tradeoff. The finding extends the individual-creativity versus collective-diversity tradeoff of \cite{doshi2024generative} in a mechanistic direction: homogenization is not a fixed property of generative AI but a tunable one, and the temperature at which the model matches the diversity of the human tradition ($T \approx 1.2$, just above the vendor default $T = 1.0$) sits immediately below the temperature at which its output stops being recognizable as belonging to the genre at all (refer to Fig.~\ref{fig:phase} for specific numbers). We note that the high-temperature regime is diverse only in a degenerate sense: pairwise similarities near zero indicate near-orthogonal vocabularies, that is, texts that share no conventions rather than texts that vary within a convention.

Secondly, the human tradition itself displays stable collective structure. The within-human baselines computed from the 2025 dataset ($n = 33$, mean similarity $\approx 0.26$) and the independently collected 2026 dataset ($n = 7$, mean $\approx 0.28$) are statistically indistinguishable, with overlapping dispersion bands. Human authors writing one year apart produce flitzes at a characteristic level of mutual resemblance. This constancy is what one expects of a functioning tradition: a shared convention that individual creativity varies around, analogous to the quenched disorder against which our spin glass model operates. Pairwise similarity heatmaps for both human-human and AI-AI comparisons are provided in Fig.~\ref{fig:hhcos} and Fig.~\ref{fig:aacos} in Appendix~B.

\subsection{Both human and AI judges demonstrate bias towards human-written flitzes.}\label{sec:r2}

\begin{figure*}[!t]%
\centering
\includegraphics[width=\textwidth]{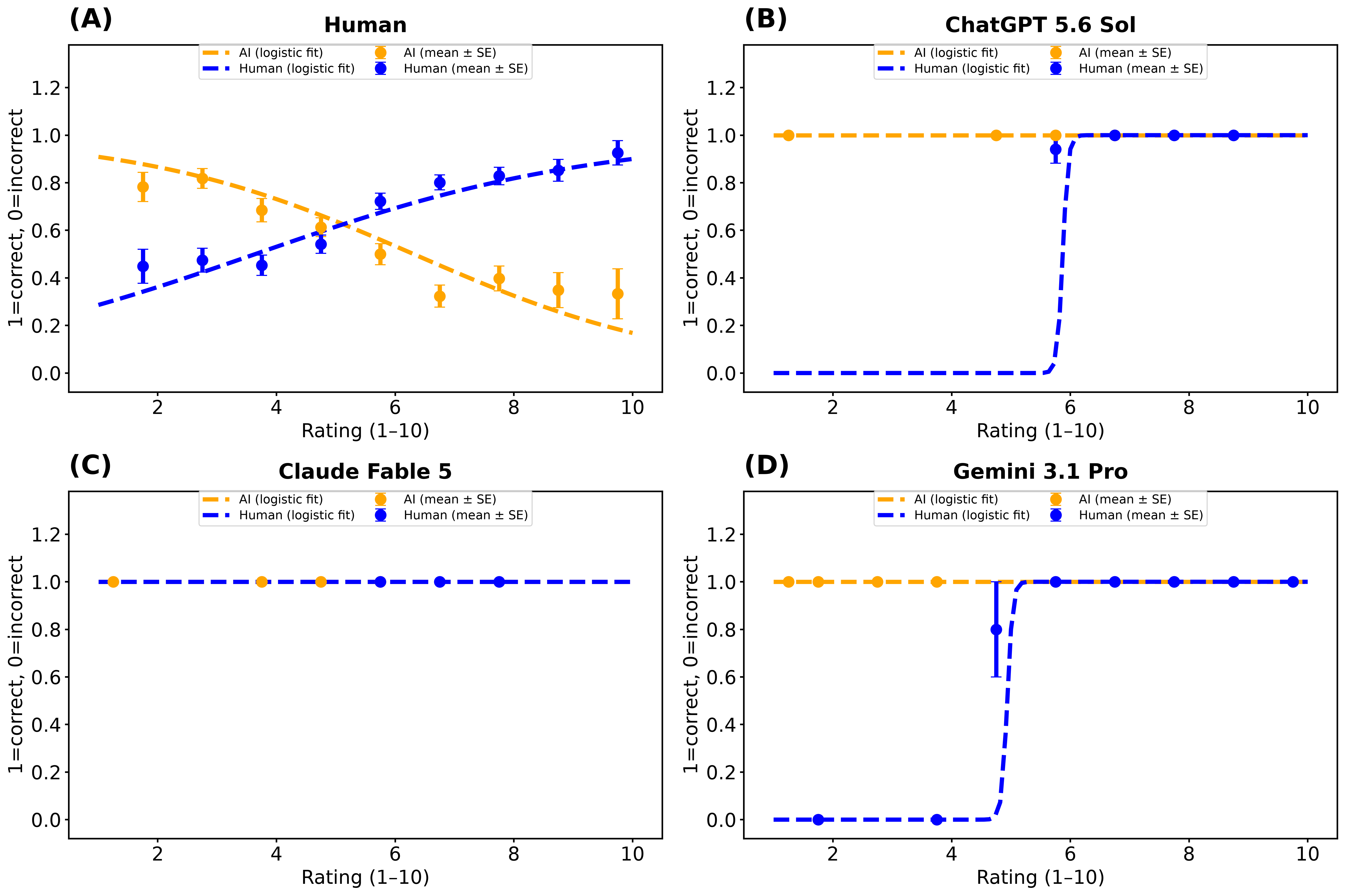}
\caption{Detection accuracy versus assigned quality rating for human and AI judges. (A) Human respondents (2025 wave, $n = 116$): mean correct-classification rate ($\pm 1$ standard error) at each rating level, shown separately for AI-generated (orange) and human-written (blue) flitzes, with logistic fits (dashed). Human judges increasingly classify highly rated flitzes as human regardless of true provenance. (B-D) The same instrument administered to ChatGPT 5.6 Sol, Claude Fable 5, and Gemini 3.1 Pro (one classification and one rating per flitz; no standard errors or fits are shown for $n = 1$ per item). Each model classifies nearly all 18 flitzes correctly in all 10 runs while assigning systematically higher ratings to human-written flitzes.}\label{fig:judges}
\end{figure*}

Our second result concerns the biases of the judges themselves, human and artificial. In the 2025 survey wave, 116 human respondents rated 9 pairs of human-written and AI-generated flitzes on a 1-10 quality scale and classified each flitz as human- or AI-authored. In 2026 we administered the identical survey to three frontier LLMs (ChatGPT 5.6 Sol, Claude Fable 5, and Gemini 3.1 Pro), each judging the same 18 flitzes. Fig.~\ref{fig:judges} shows classification accuracy as a function of the rating the judge assigned.

For human judges (Fig.~\ref{fig:judges}A), perceived quality functions as a humanness heuristic. The probability of classifying a flitz as human-written increases logistically with its rating: human-written flitzes receiving a rating of 9-10 are correctly attributed roughly 90\% of the time, while human-written flitzes rated 1-2 are misattributed to AI more often than not. The mirror-image trend holds for AI-generated flitzes. If a flitz sounds good to a human respondent, the respondent concludes a human wrote it. This replicates, inside a courtship genre, the flawed-heuristics finding of \cite{jakesch2023human} and the quality-humanness inversion documented for poetry \cite{porter2024ai}.

Across ten independent runs per model (180 judgements each, 540 in total), all three models detect provenance almost perfectly: Claude Fable 5 classifies every flitz correctly in every run (180/180), ChatGPT 5.6 Sol mistakenly classifies once (179/180, 99.4\%), and Gemini 3.1 Pro makes false classifications seven times (173/180, 96.1\%), for an aggregate accuracy of 532/540 (98.5\%), far above the 72.5\% pooled accuracy of the 2026 human respondents on the same items. The errors are strikingly one-sided: all eight are human-written flitzes labeled as AI, confined to just two items (flitzes number 8 and 12), and no model ever labeled an AI-generated text as human-written. Yet despite recognizing AI-generated text as AI-generated text, all three models assign systematically higher quality ratings to the human-written flitzes than to the AI-generated ones (mean ratings 6.9 vs.\ 3.9 for Claude, 7.2 vs.\ 4.3 for ChatGPT, and 7.3 vs.\ 2.2 for Gemini). The size of the gap differs by model: Gemini rates every AI flitz at 4 or below, whereas Claude and ChatGPT place most AI flitzes at mid-scale (5-7) and reserve the floor rating of 1 for the same three items (flitzes number 3, 17, and 18) in all thirty runs. AI judges therefore dissociate provenance from quality: they are more reliable detectors that nevertheless prefer human writing. Because per-rating accuracy is at or near 1.0 across almost the entire scale, the logistic fits in Figs.~1(B)-(D) are essentially degenerate, and the apparent drops to 0.0 in Fig.~1(D) reflect only the handful of Gemini runs in which number 8 and 12 were misjudged, all of which fall at low ratings. These misjudgments are themselves informative. On number 12, a human-written flitz, Gemini's verdict varied across runs, and its rating tracked the verdict: in the seven runs it judged the flitz human-written it rated it 5-7 (mean 5.7), while in the three runs it judged it AI-generated it rated it 2-4 (mean 2.7), including the single rating of 2 visible in the figure; number 8 shows the same pattern (mean 6.7 as human-written'' versus 4.3 as AI-generated''). The same text thus receives a different quality score depending on the judge's belief about its origin, a within-item version of the quality-humanness heuristic that drives human errors. We note this tendency tentatively, since it rests on two items and a single model, but it motivates a research direction on the mechanisms under which AI rates AI-generated texts as inferior to human-written texts, even if the texts are not obviously AI-generated to humans. Possible mechanisms include preference-tuning on human judgments that encode the human quality-humanness heuristic, and stylometric regularities of AI-generated text that are salient to AI systems but not to people. The asymmetry matters practically because AI systems increasingly moderate and contribute to human communication, and their failure modes evidently differ from ours.

\subsection{Temperature-induced phase transitions are consistent across data, text mining, and mathematical model simulation.}\label{sec:r3}

\begin{figure*}[!t]%
\centering
\includegraphics[width=\textwidth]{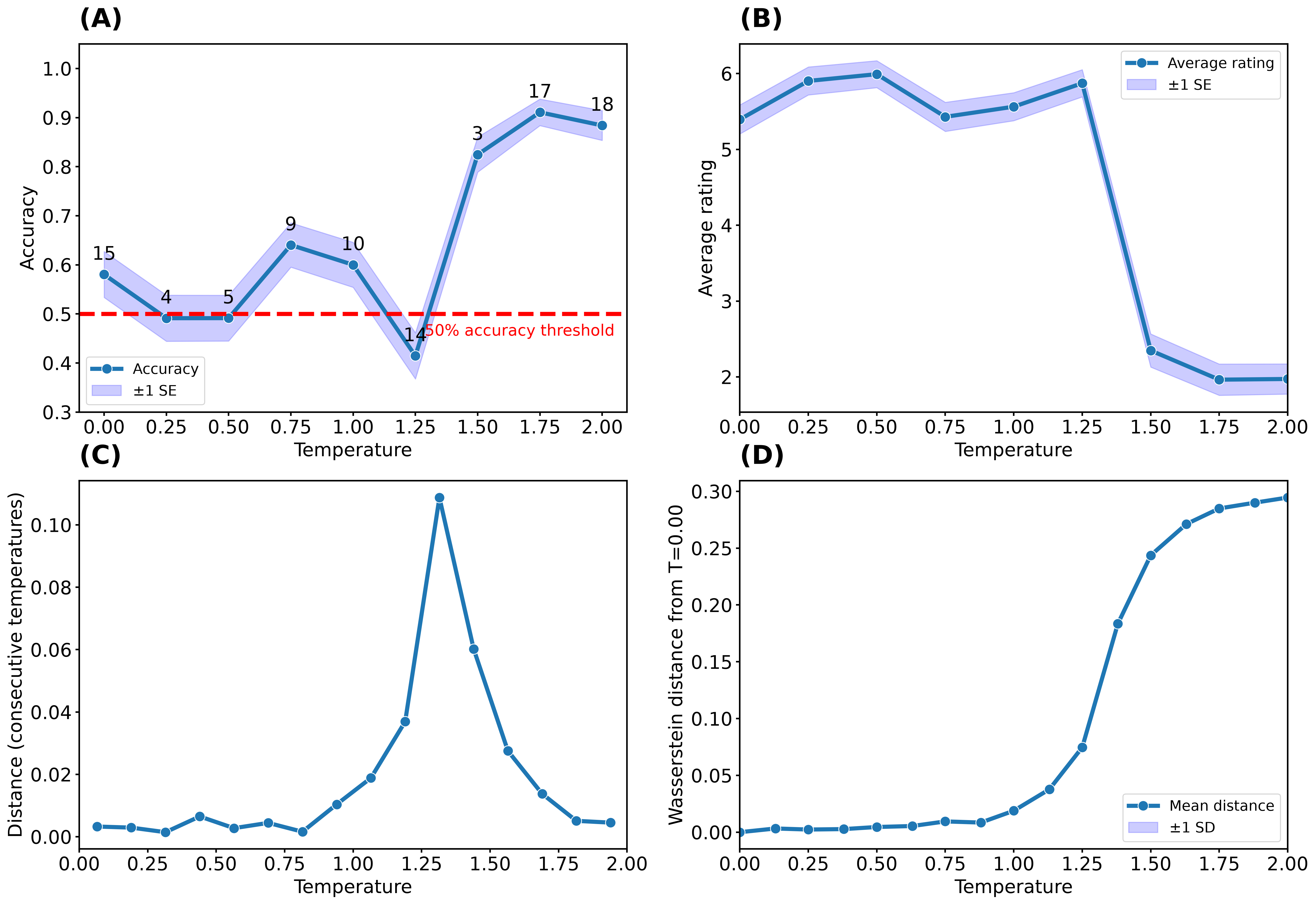}
\caption{Consistency in phase transitions between biased and unbiased data (2025 wave). (A) Accuracy in identifying human-written versus AI-generated flitzes across temperatures, with the 50\% unbiased-coin threshold marked; annotations give the number of respondents seeing each pair. Accuracy is near chance for $T \in [0, 1.00)$, reaches a local minimum of 41\% at $T = 1.25$, and rises to 91\% at $T = 1.75$. (B) Mean quality rating of AI-generated flitzes (1-10 scale, shading $\pm 1$ standard deviation): ratings are stable near 5.4-6.0 through $T = 1.25$ and collapse to approximately 2 beyond it. (C) Wasserstein distance between cosine-similarity distributions at consecutive temperatures, peaking sharply between $T = 1.25$ and $T = 1.38$. (D) Wasserstein distance from the $T = 0$ baseline: flat below $T = 1.00$, a pronounced sigmoidal jump at $T \approx 1.38$, and a plateau beyond $T \approx 1.63$.}\label{fig:phase}
\end{figure*}

\begin{figure}[!t]%
\centering
\includegraphics[width=\columnwidth]{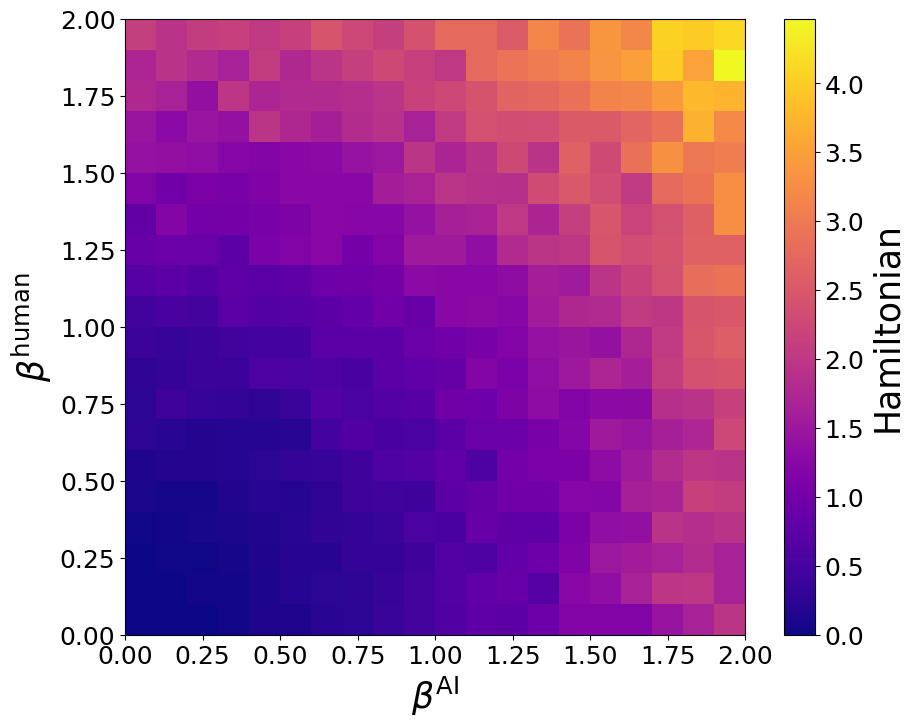}
\caption{Phase diagram of the mathematical spin glass model: average energy as a function of the inverse temperatures $\beta^{\mathrm{human}}$ and $\beta^{\mathrm{AI}}$, computed by Metropolis-Hastings simulation with burn-in at each grid point.}\label{fig:phasediagram}
\end{figure}

Our third result establishes that the phase transition identified in the diversity analysis appears consistently in three independent places: in biased human survey data, in unbiased information-retrieval statistics of the text corpus, and in the mathematical model. This result is summarized in Fig.~\ref{fig:phase} and Fig.~\ref{fig:phasediagram}.

In the survey data (Fig.~\ref{fig:phase}(A)), detection accuracy is statistically indistinguishable from an unbiased coin for $T \in [0, 1.00)$ (49-64\%), dips to a local minimum of 41\% at $T = 1.25$, where AI-generated flitzes are actively mistaken for human ones, and then jumps to 82-91\% for $T \geq 1.5$. Perceived quality (Fig.~\ref{fig:phase}(B)) follows the same pattern: mean ratings hold between 5.4 and 6.0 up to $T = 1.25$, with a local maximum at $T = 0.5$ and a secondary peak at $T = 1.25$, then collapse to approximately 2 at higher temperatures as the output becomes erratic and incoherent. The unbiased text statistics locate the transition in the same window without reference to any human judgment. The Wasserstein distance between cosine-similarity distributions at consecutive temperatures (Fig.~\ref{fig:phase}(C)) hovers near zero through $T \approx 1.00$ and spikes by an order of magnitude between $T = 1.25$ and $T = 1.38$; the distance from the $T = 0$ reference distribution (Fig.~\ref{fig:phase}(D)) is sigmoidal with its jump at $T \approx 1.38$ and a plateau beyond $T \approx 1.63$, indicating that the high-temperature distributions are far from all low-temperature ones and stable among themselves. On the model side, simulation of the coupled multi-species spin glass system produces an average-energy phase diagram over the $(\beta^{\mathrm{human}}, \beta^{\mathrm{AI}})$ plane with distinct low- and high-energy regimes separated by a curved boundary (Fig.~\ref{fig:phasediagram}), consistent with an SK-type phase transition in the temperature direction \cite{panchenko2012sherrington,amit1985spin}. The three methodologies, which share no data, agree on a transition window $T \in [1.25, 1.38]$, slightly above the vendor default temperature $T = 1.00$.

Fig.~\ref{fig:phasediagram} shows the average energy of the coupled three-species system as a function of $(\beta^{\mathrm{human}}, \beta^{\mathrm{AI}})$, each ranging over $[0, 2]$. Each point is the mean energy over Metropolis-Hastings updates following a burn-in period. The diagram exhibits distinct low-energy and high-energy regimes separated by a curved (approximately quadratic) boundary, the model-side signature of the phase transition: at higher $\beta$ (lower temperature), configurations in which people align with both human-written and AI-generated flitzes dominate, while at lower $\beta$ (higher temperature) the system disorders.

\subsection{Clustering is consistent between data and mathematical model simulation.}\label{sec:r4}

\begin{figure}[!t]%
\centering
\includegraphics[width=\columnwidth]{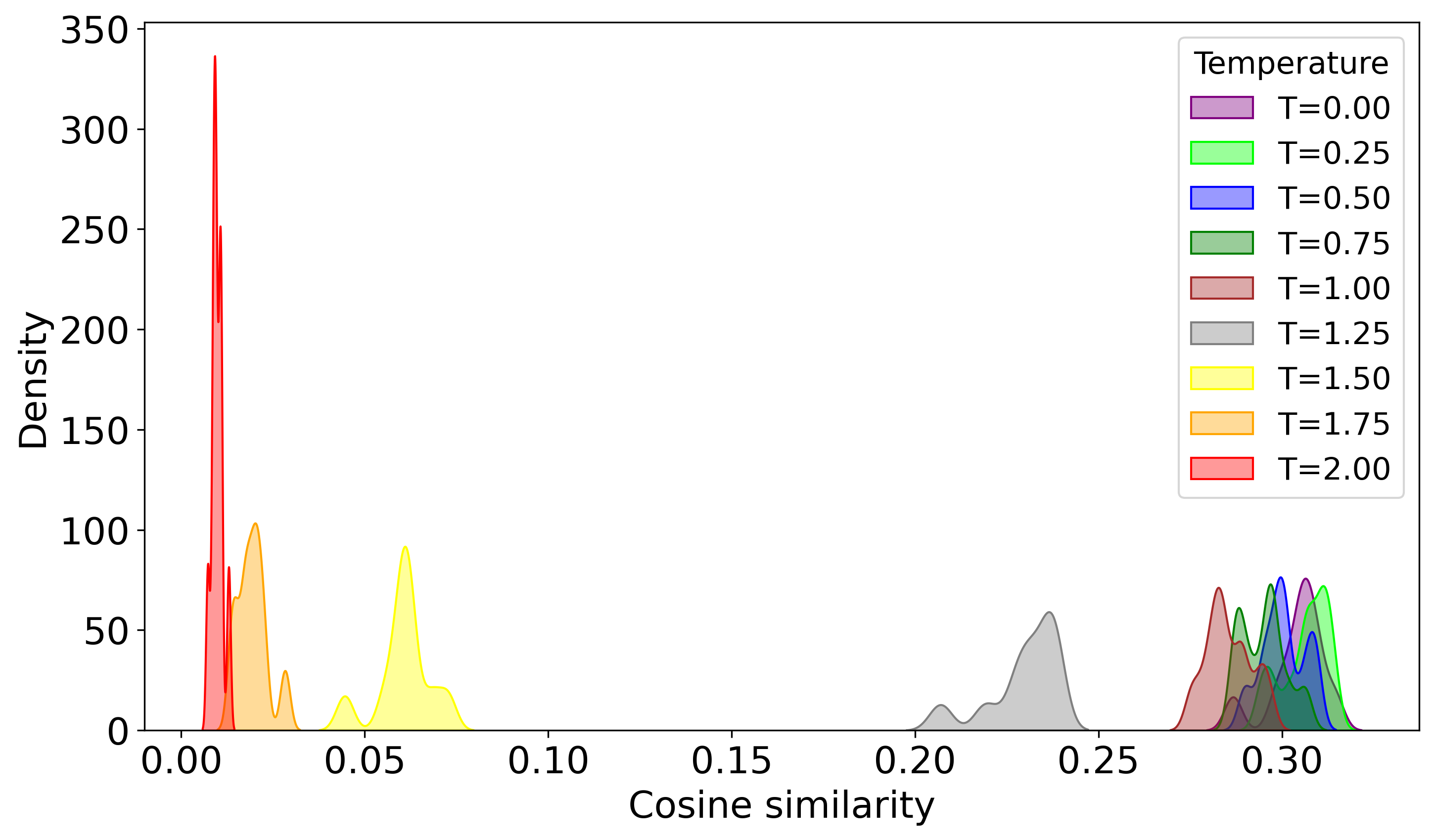}
\caption{Clustering phenomenon in the data. Kernel density estimates of cosine-similarity scores between the 33 human-written flitzes and their paired AI-generated flitzes at nine representative temperatures. Cold-color distributions ($T \leq 1.00$) and hot-color distributions ($T \geq 1.5$) form two distinct clusters, with the $T = 1.25$ distribution (gray) occupying a transitional position.}\label{fig:kde}
\end{figure}

Our fourth result concerns the structure on either side of the transition. Fig.~\ref{fig:kde} shows kernel density estimates (KDE) of the cosine-similarity scores between human-written flitzes and their AI-generated counterparts at each temperature. The distributions form two well-separated clusters: a low-temperature cluster ($T \in [0, 1.00]$) concentrated near similarity $0.28$-$0.31$, and a high-temperature cluster ($T \geq 1.5$) concentrated below $0.08$, with the distribution at $T = 1.25$ isolated between them near $0.21$-$0.24$ as a transitional state. Increased temperature beyond the transition produces diminishing structural change, mirroring the plateau in Fig.~\ref{fig:phase}(D).

\begin{figure*}[!t]%
\centering
\includegraphics[width=\textwidth]{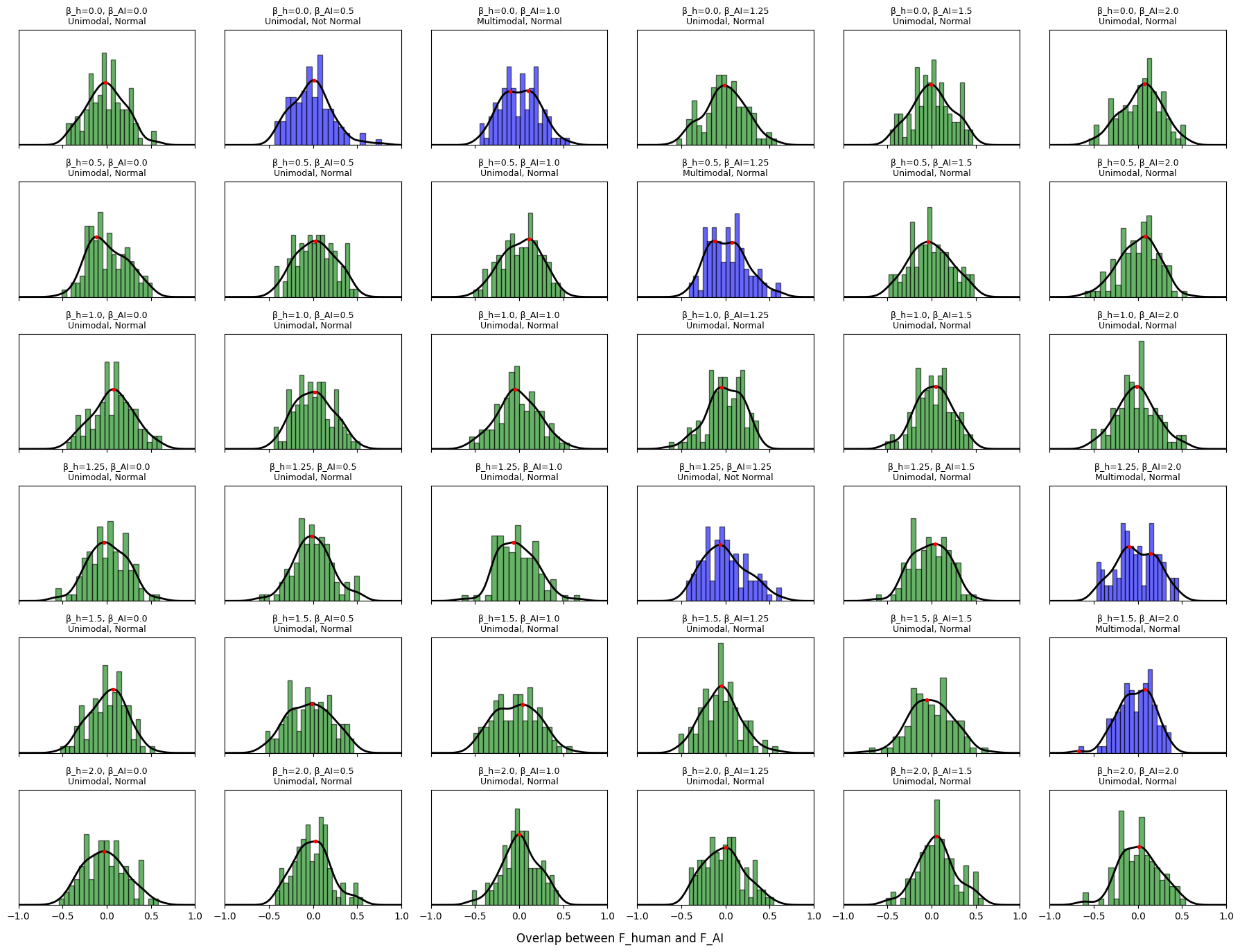}
\caption{Clustering in the mathematical spin glass model. Distributions of the overlap between human-flitz and AI-flitz representations for 36 combinations of $(\beta^{\mathrm{human}}, \beta^{\mathrm{AI}})$, with rows indexing $\beta^{\mathrm{human}}$ and columns $\beta^{\mathrm{AI}}$. Histograms are fitted with kernel density estimates; green panels are unimodal and normal (Shapiro-Wilk), blue panels are multimodal or non-normal. Multimodality occurs at $(0, 1), (0.5, 1.25), (1.25, 2)$, and $(1.5, 2)$; the $(1.25, 1.25)$ panel is unimodal but non-normal.}\label{fig:modelclust}
\end{figure*}

The spin glass model reproduces this clustering. Fig.~\ref{fig:modelclust} shows simulated distributions of the overlap between the human-flitz and AI-flitz representations for 36 combinations of inverse temperatures $(\beta^{\mathrm{human}}, \beta^{\mathrm{AI}}) \in \{0, 0.5, 1.00, 1.25, 1.5, 2.00\}^2$. Most combinations yield unimodal, Gaussian overlap distributions, the signature of a single dominant alignment state. Multimodal distributions occur only at $(\beta^{\mathrm{human}}, \beta^{\mathrm{AI}}) = (0, 1), (0.5, 1.25), (1.25, 2)$, and $(1.5, 2)$, and the distribution at $(1.25, 1.25)$ is unimodal but significantly non-normal. Multimodality concentrates along the $\beta^{\mathrm{AI}}$ direction at values adjacent to the empirical transition, consistent with metastable competing states near criticality in self-attention dynamics \cite{geshkovski2024dynamic} and with the phase transition behavior in the data.

\subsection{Findings are robust across space but not time.}\label{sec:r5}

\begin{figure}[!t]%
\centering
\includegraphics[width=\columnwidth]{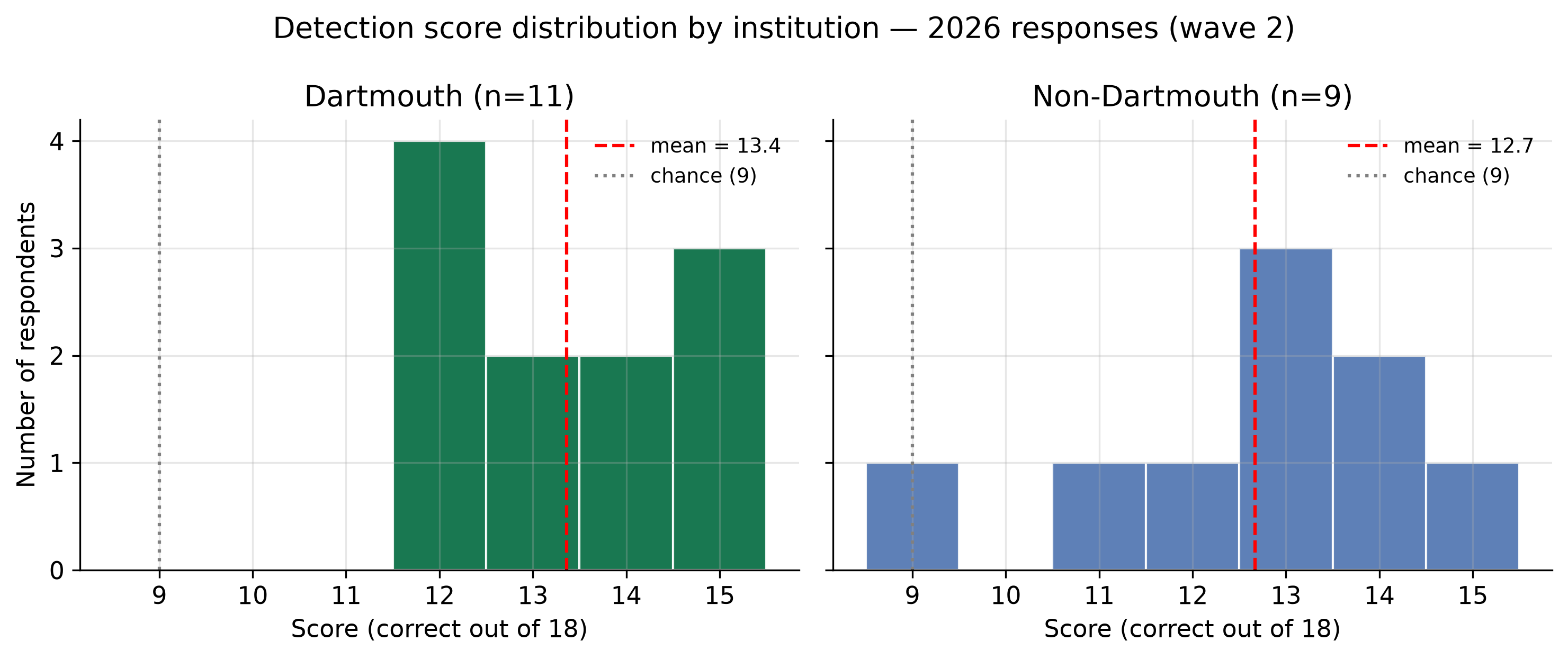}
\caption{Detection score distributions by institution in the 2026 wave. Histograms of per-respondent scores (correct classifications out of 18) for Dartmouth-affiliated ($n = 11$, mean 13.4) and non-Dartmouth ($n = 9$, mean 12.7) respondents; the dotted line marks the chance score of 9. Both groups sit well above chance and do not differ significantly from each other.}\label{fig:institution}
\end{figure}

\begin{figure}[!t]%
\centering
\includegraphics[width=\columnwidth]{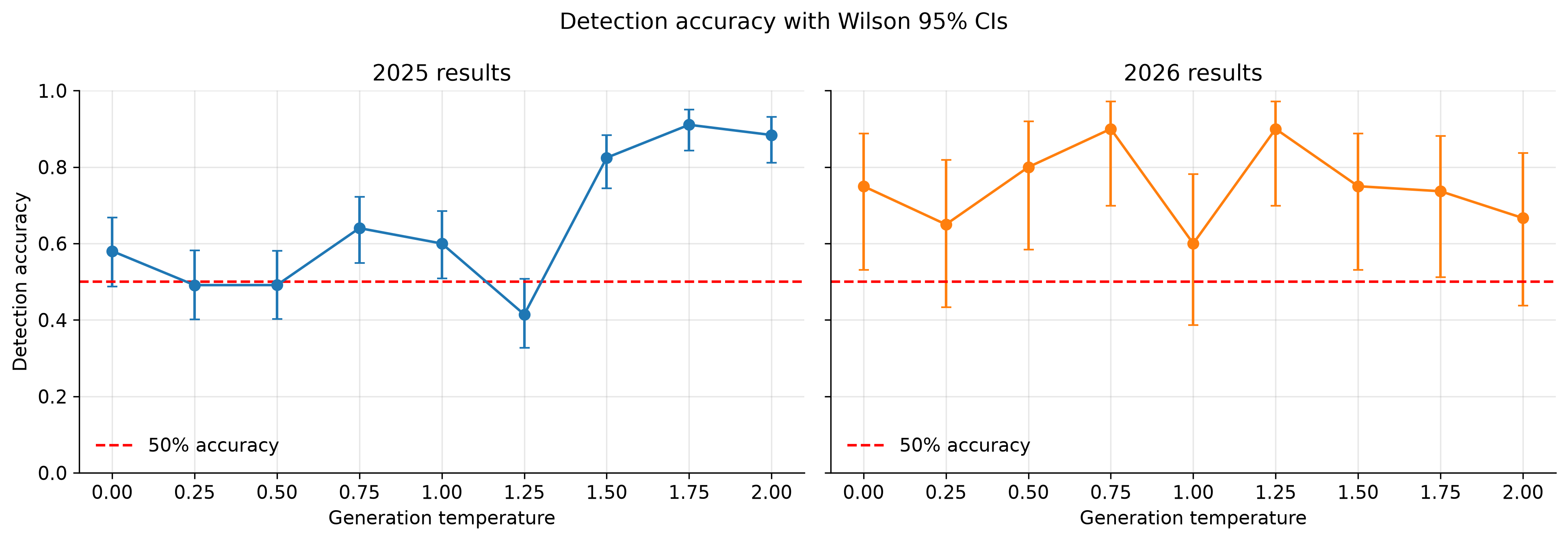}
\caption{Detection accuracy by generation temperature in the 2025 wave (left, $n = 116$) and the 2026 wave (right, $n = 20$), with Wilson 95\% confidence intervals and the 50\% chance line. The 2025 wave shows the chance-level regime at low temperature, the local minimum at $T = 1.25$, and the high-accuracy regime beyond; in the 2026 wave accuracy is above chance at every temperature, including the low-temperature regime in which 2025 respondents were at chance.}\label{fig:waves}
\end{figure}

Our fifth result tests whether these findings depend on membership in the culture that produced the genre, and whether they are stable in time. In 2026 we administered the original 18-item survey to a second wave of respondents ($n = 20$), recording whether each respondent was affiliated with Dartmouth ($n = 11$) or not ($n = 9$). Fig.~\ref{fig:institution} shows the distribution of detection scores by institutional affiliation, and Fig.~\ref{fig:waves} compares detection accuracy by temperature across the two waves.

Spatially, detection does not require cultural membership. Dartmouth respondents scored a mean of $13.4 \pm 1.3$ out of 18 (74.2\% accuracy) and non-Dartmouth respondents $12.7 \pm 1.8$ (70.4\%); the difference is not significant (Welch's $t \approx 0.97$, $\mathrm{df} \approx 14$, $p \approx 0.35$; Mann-Whitney $p \approx 0.51$), while each group individually exceeds chance (exact binomial $p < 10^{-6}$ for both). Insider knowledge of flitzing conventions confers, at most, a small advantage that our sample cannot resolve. The signal that separates AI-generated from human-written flitzes is therefore carried by properties of the text that travel beyond the local culture, consistent with the temperature mechanism rather than with unique cultural knowledge.

Temporally, the picture drifted between waves in one specific way. The 2026 respondents are above chance at every temperature (60-90\%), including the low-temperature regime $T \leq 1.00$ in which the much larger 2025 sample was statistically at chance, and the 2025 local minimum at $T = 1.25$ (41\%) is replaced in 2026 by accuracy near 90\% at the same temperature. The transition structure survives in attenuated form (accuracy still varies with temperature, and the smaller 2026 sample carries wide confidence intervals), but the  classification difficulty at low temperature that defined the 2025 results has eroded. We interpret this cautiously, because the waves differ in size and are not a panel of the same individuals. The pattern is, however, exactly what population-level adaptation predicts: between the two waves, LLM-inflected style became a familiar object of everyday reading \cite{kobak2025delving,yakura2024empirical}, and plausibly, respondents can differentiate it more easily. The Turing window, in this genre, appears to be closing from the human side for legacy model outputs: some adaptation at least for ChatGPT-4o can be inferred cautiously.

\section{Conclusion and discussion}

When a generative model is embedded in a human social network, our results indicate that it does not perturb the culture uniformly. Instead, its participation is organized by a control parameter into two regimes separated by a sharp transition at critical temperature $T^{*} \in [1.25, 1.38]$. Below $T^{*}$ the model is conformist. Its output passes as human (2025 detection at chance, and actively mistaken for human at $T = 1.25$), is rated as highly as human writing, and is more homogeneous than the tradition it imitates, with within-AI similarity roughly 50\% above the within-human baseline. Above $T^{*}$ the model is deviant: instantly detected, poorly rated, and incoherent, sharing conventions neither with the tradition nor with its own other outputs. On the other hand, the human tradition holds its collective structure fixed across years. The near-term cultural risk suggested by these results is therefore not disruption from without but homogenization from within: at deployment-relevant temperatures, the model participates fluently while compressing the diversity that gives a courtship genre its signaling value.

This mapping connects directly to constructs in cultural psychology. Tightness-looseness theory characterizes cultures by the strength of their norms and their tolerance of deviance \cite{gelfand2011differences}; the flitz genre is a tight micro-culture, with explicit formal conventions and strong sanctions of ridicule for violating them. In our framework the sampling temperature is a continuous knob that moves a single agent from maximal norm adherence to maximal deviance, and the reception data locate the culture's tolerance threshold: deviations below $T^{*}$ are absorbed, deviations above it are detected and penalized in ratings. The spin glass model makes the same point structurally. The human tradition supplies quenched disorder, a fixed random landscape of conventions against which new texts are evaluated, and one generative model serving many senders replaces independent draws of that disorder with correlated ones, the condition under which algorithmic monoculture degrades collective outcomes even while improving individual ones \cite{kleinberg2021algorithmic,doshi2024generative}. In cultural-evolutionary terms, temperature acts as the machine analog of the mutation rate in a system otherwise dominated by conformist transmission \cite{brinkmann2023machine}: too low, and the population of texts loses variation; too high, and the variants fall outside the tradition altogether.

The Turing experiment adds a second, less anticipated asymmetry. Human judges use perceived quality as a humanness heuristic and are exploitable by exactly the conventional, competent text that low-temperature generation produces \cite{jakesch2023human,porter2024ai}. AI judges are not exploitable this way: they detect provenance almost perfectly in this genre, yet they assign human writing higher quality, including relative to output of their own model family. Detection and evaluation, which covary in humans, dissociate in machines. One implication is methodological: LLM-as-judge pipelines are not simply cheap simulations of human raters, because their bias structure differs in kind, not merely in degree \cite{argyle2023out}. Another is sociotechnical: if machine gatekeepers (spam filters, authenticity classifiers, ranking systems) become the effective audience for courtship and other relational text, the selection pressures on that text will differ from the pressures human audiences impose, and Goodhart dynamics on ``sounding human'' may follow. The mechanism behind the AIs' preference for human writing deserves dedicated study; preference-tuning on human judgments and stylometric self-recognition are both consistent with our data.

The robustness results discipline the interpretation in two directions. Spatial robustness (Dartmouth versus non-Dartmouth respondents statistically indistinguishable, both above chance) implies that the detectability signal is carried by temperature-dependent properties of the text itself rather than by insider cultural knowledge, which supports reading the transition as a property of the generative process rather than of the local audience. Temporal drift, in which the 2026 wave detects low-temperature AI text above chance where the 2025 wave could not, implies that the human side of the loop is not stationary. Language communities adapt to the AI systems that write among them \cite{saadat2024impact,yakura2024empirical,kobak2025delving}, and our two waves bracket a period of rapid public exposure to LLM prose. A feedback loop is therefore plausible in both directions: low-temperature generation nudges the tradition toward the model's median style even as readers grow more sensitive to that style. Modeling this co-adaptation, for instance by letting $\beta^{\mathrm{human}}$ and the coupling matrices evolve in time, is a natural extension of the framework, and the rigorous multi-species SK literature \cite{bates2019replica,batesmulti2022,collins2024free,collins2025order} provides the analytic tools such an extension would draw on.

Because Dartmouth College is a hyper-WEIRD population \cite{henrich2010weirdest, dartmouth2024class2028, usnews2026dartmouth, dartmouth2025class2029profile} and the flitz sits near the center of a frontier model's training distribution \cite{atari2023which,tao2024cultural}, our measurements plausibly constitute an upper bound for the model's performance. Where GPT-4o can homogenize a tradition it knows intimately, models should flatten more distant genres and sociolects at least as strongly, a concern already documented in the covert penalization of for instance non-Caucasian demographics~\cite{hofmann2024ai} and the underperformance of models on culturally grounded scenarios \cite{shi2024culturebank}. Our contribution to that literature is a complementary method: rather than injecting cultural knowledge into models or benchmarking them against static test sets, one can measure how faithfully a model inhabits one living practice, at what diversity cost, and with what detectability, as a function of an interpretable control parameter.

Several limitations bound these conclusions. Flitzes are culturally idiosyncratic, and although our respondents span institutions, the genre itself may not generalize to other forms of human-AI linguistic interaction. Temperature is the only generation parameter we vary; interactions with nucleus or min-$p$ sampling \cite{nguyen2024turning,holtzman2020curious} could reshape the transition. The Flitz generation is recursive (ChatGPT-4o summarizes each human flitz into a profile from which ChatGPT-4o then generates), which may constrain stylistic range, particularly at low temperature, and all generated flitzes come from a single model, although the judge experiment spans three model families. The 2026 wave is small ($n = 20$), its respondents differ from the 2025 respondents, and the AI judges only represent the current state of development, judging AI-written texts from legacy models. Finally, our spin glass model is a minimal mechanism: it reproduces the transition and the clustering qualitatively but is not fitted to the text data, and the correspondence between its inverse temperatures and the LLM's sampling temperature is structural rather than calibrated.

These limitations inspires future work. Replication in other codified genres (sonnets, wedding toasts, cover letters, workplace emails) would test whether tight micro-genres generically exhibit a conformist-deviant transition near the vendor default temperature. Panel designs following the same respondents across years would turn our temporal-drift observation into a measurement of human adaptation rates, and letting model output feed back into the training or prompting loop would probe the homogenization dynamics directly. On the mathematical side, fitting the coupled SK model's couplings to data, and analyzing the crossing of the detection and adoption curves as a two-parameter phase diagram, would move the model from consistency argument toward prediction. As generative systems continue to write within human traditions, from love letters to liturgy, frameworks that quantify when they conform, when they deviate, and what their participation does to collective diversity will be necessary to keep both the AI systems and the cultures legible. Analyzing these phase transitions is a step toward AI systems whose participation in human culture is interpretable, tunable, and aligned with the values of the communities they write for.

\section{Materials and methods}

\subsection{Research design and data generation}\label{sec:design}

\begin{figure}[!t]%
\centering
\includegraphics[width=\columnwidth]{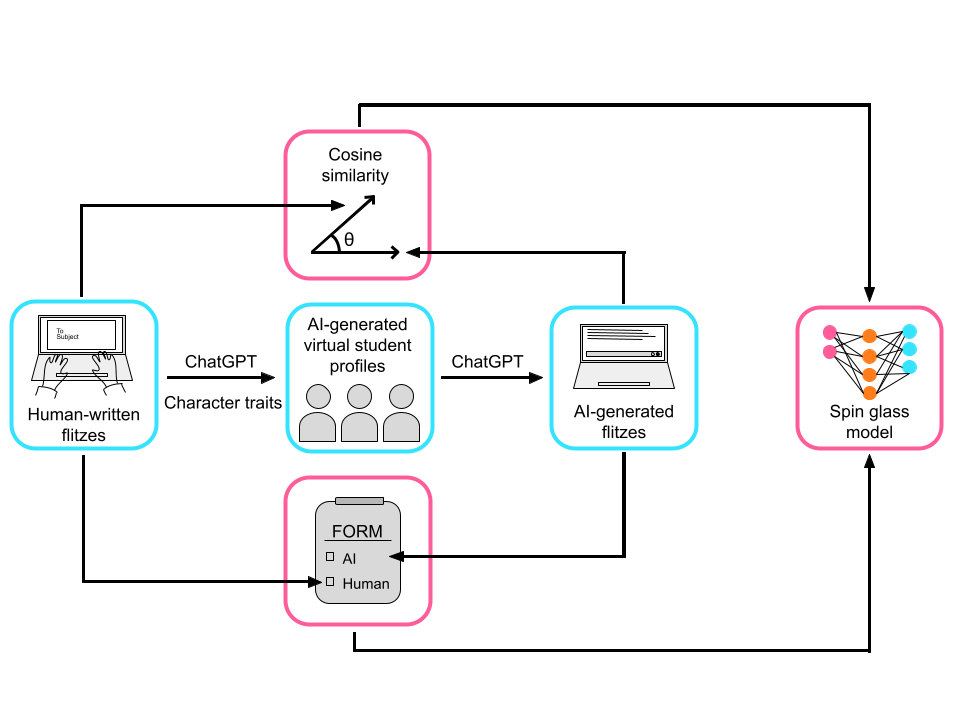}
\caption{Schematic diagram of the research design and data generation. Blue boxes indicate datasets and pink boxes indicate methods. Human-written flitzes seed AI-generated virtual student profiles, which seed AI-generated flitzes across 17 temperatures; cosine similarity compares the corpora; surveys collect human (and, in 2026, machine) classifications and quality ratings; and a coupled spin glass model characterizes the phase transition behavior.}\label{fig:schematic}
\end{figure}

Fig.~\ref{fig:schematic} summarizes our research design. We use three methodologies: (M1) statistical analysis of human and machine evaluation surveys, (M2) information retrieval and cosine similarity on collected and generated text, and (M3) mathematical modeling and simulation. These operate on four datasets: (D1) survey data on human perception in two waves, (D2) human-written flitzes collected from Dartmouth undergraduates, (D3) AI-generated flitzes produced from virtual student profiles, and (D4) AI-judge responses from three frontier LLMs.

\emph{Human-written texts.} We collected 33 human-written flitzes from 33 Dartmouth undergraduates, each used in a real social setting, in 2025. In 2026 we collected 7 additional human-written flitzes by the same procedure; these enter the analysis only as the independent 2026 within-human baseline in Fig.~\ref{fig:homog}.

\emph{AI-generated texts.} Each 2025 flitz was input to ChatGPT-4o at the default temperature $T = 1.00$ to create a corresponding virtual student profile describing tone, humor, flirtation style, and language quirks. For each of the 33 profiles, ChatGPT-4o then produced 10 independent flitzes at each of 17 temperature values $\{0, 0.125, 0.25, \ldots, 1.875, 2\}$ (rounded in figures to $\{0, 0.13, \ldots, 1.88, 2\}$), for $33 \times 10 \times 17 = 5{,}610$ AI-generated flitzes. The base prompt asked the model to read two published descriptions of flitzing and to write a flitz acting as a student with the given profile. Generation was capped at 500 tokens; all parameters other than temperature (top-$p$, frequency penalty) were held constant.

\emph{Survey waves.} The survey presents 9 pairs of human-written and AI-generated flitzes, one pair for each temperature in $\{0, 0.25, 0.5, 0.75, 1.00, 1.25, 1.50, 1.75, 2.00\}$, and asks respondents to rate each flitz from 1 (worst) to 10 (best) and to classify it as human- or AI-authored. Wave 1 (2025) collected 117 respondents from inside and outside Dartmouth; one respondent who answered a single question was excluded, leaving $n = 116$. Wave 2 (2026) administered the identical survey to $n = 20$ new respondents and recorded institutional affiliation (11 Dartmouth, 9 non-Dartmouth).

\emph{AI judges.} In 2026 we presented the same 18 flitzes, in the survey order, to three frontier models: ChatGPT 5.6 Sol \cite{chatgpt56sol}, Claude Fable 5 \cite{claudefable5}, and Gemini 3.1 Pro \cite{gemini31pro}. Each model received the same instructions as human respondents (rate 1-10; classify human versus AI) and produced one rating and one classification per flitz, repeating this process 10 times with each model.

\subsection{Information retrieval with cosine similarity and Wasserstein distance}\label{sec:ir}

We vectorize flitzes with term frequency-inverse document frequency (TF-IDF) \cite{tfidf-scikit}, in which the score of term $t$ in document $d$ is
\begin{equation}
\mathrm{TF\mbox{-}IDF}(t, d) = \left(1 + \log(\mathrm{tf}_{t,d})\right) \cdot \left(\log \frac{1 + N}{1 + \mathrm{df}_t} + 1\right),
\label{eq:tfidf}
\end{equation}
where $\mathrm{tf}_{t,d}$ is the raw count of $t$ in $d$, $\mathrm{df}_t$ the number of documents containing $t$, and $N$ the size of the collection of texts. Similarity between two TF-IDF vectors $\mathbf{u}$ and $\mathbf{v}$ is the cosine
\begin{equation}
\mathrm{cos}(\mathbf{u}, \mathbf{v}) = \frac{\mathbf{u} \cdot \mathbf{v}}{\lVert \mathbf{u} \rVert \, \lVert \mathbf{v} \rVert} \in [0, 1],
\label{eq:cosine}
\end{equation}
nonnegative here because TF-IDF vectors are nonnegative (\texttt{TfidfVectorizer} in scikit-learn). Three similarity statistics appear in the Results: (i) paired human-AI similarity, computed between each human-written flitz and each AI-generated flitz derived from its profile, whose distributions across pairs give Fig.~\ref{fig:kde}; (ii) within-AI cross-profile similarity at temperature $T$, the mean of $\mathrm{cosine}$ over pairs of AI flitzes generated from different profiles at that temperature (Fig.~\ref{fig:homog}); and (iii) within-human similarity, the mean over all pairs of the 33 (respectively 7) human flitzes, providing the 2025 and 2026 baselines in Fig.~\ref{fig:homog}.

To locate structural change across temperature we use the Wasserstein distance \cite{villani2008optimal,piccoli2016properties} between cosine-similarity distributions. For distributions $\mu$ and $\nu$ on a metric space $\mathcal{X}$,
\begin{equation}
W_p(\mu, \nu) = \left( \inf_{\gamma \in \Gamma(\mu, \nu)} \int_{\mathcal{X} \times \mathcal{X}} d(x, y)^p \, d\gamma(x, y) \right)^{1/p},
\label{eq:wasserstein}
\end{equation}
with $\Gamma(\mu, \nu)$ the set of couplings; we use $p = 1$. We compute $W_1$ between consecutive temperatures (Fig.~\ref{fig:phase}(C)) and between each temperature and the $T = 0$ baseline (Fig.~\ref{fig:phase}(D)). Unlike divergences that ignore geometry, $W_1$ is a smooth, metric-aware functional of the full distributions.

\subsection{Spin glass model and simulation}\label{sec:model}

We model the human-AI interaction network as two coupled multi-species bipartite spherical SK models \cite{panchenko2012sherrington,bates2019replica,batesmulti2022,collins2024free,collins2025order}; for general background on spin glasses see \cite{talagrand2010mean,talagrand2011mean,bovier2006statistical,mezard1987spin,mezard2023spin}. In a bipartite structure, interactions occur only between, not within, species, a structure natural for people interacting with texts and used widely for complex systems \cite{pavlopoulos2018bipartite,he2016birank,huynh2025inferring}. A bipartite multi-species spherical SK model consists of two groups of spins
\begin{equation}
\boldsymbol{\sigma} \in S^{n-1}, \qquad \boldsymbol{\tau} \in S^{m-1},
\label{eq:spins}
\end{equation}
where $S^{d-1} = \{ \mathbf{x} \in \mathbb{R}^d : \lVert \mathbf{x} \rVert^2 = d \}$ is the radius-$\sqrt{d}$ sphere, with Hamiltonian
\begin{equation}
H(\boldsymbol{\sigma}, \boldsymbol{\tau}) = \frac{1}{\sqrt{n + m}} \sum_{i=1}^{n} \sum_{j=1}^{m} J_{ij} \sigma_i \tau_j, \qquad J_{ij} \sim \mathcal{N}(0, 1),
\label{eq:hamiltonian}
\end{equation}
and Gibbs measure $p(\boldsymbol{\sigma}, \boldsymbol{\tau}) = Z^{-1} e^{\beta H(\boldsymbol{\sigma}, \boldsymbol{\tau})}$ at inverse temperature $\beta > 0$, where $Z$ normalizes over the product of spheres.

We define three species: people $\mathbf{X} \in S^{n_{\mathrm{people}} - 1}$, human-written flitzes $\mathbf{F}^{\mathrm{human}} \in S^{n_{\mathrm{fh}} - 1}$, and AI-generated flitzes $\mathbf{F}^{\mathrm{AI}} \in S^{n_{\mathrm{fAI}} - 1}$, coupled through two bipartite Hamiltonians
\begin{align}
H(\mathbf{X}, \mathbf{F}^{\mathrm{human}}) &= \frac{1}{\sqrt{n_{\mathrm{people}} + n_{\mathrm{fh}}}} \sum_{i=1}^{n_{\mathrm{people}}} \sum_{j=1}^{n_{\mathrm{fh}}} J^{\mathrm{human}}_{ij} X_i F^{\mathrm{human}}_j, \nonumber \\
H(\mathbf{X}, \mathbf{F}^{\mathrm{AI}}) &= \frac{1}{\sqrt{n_{\mathrm{people}} + n_{\mathrm{fAI}}}} \sum_{i=1}^{n_{\mathrm{people}}} \sum_{j=1}^{n_{\mathrm{fAI}}} J^{\mathrm{AI}}_{ij} X_i F^{\mathrm{AI}}_j,
\label{eq:coupled}
\end{align}
with independent standard Gaussian couplings and associated Gibbs measures at inverse temperatures $\beta^{\mathrm{human}}$ and $\beta^{\mathrm{AI}}$, respectively. The two subsystems share the species $\mathbf{X}$, which couples them.

We approximate the joint distribution over $(\mathbf{X}, \mathbf{F}^{\mathrm{human}}, \mathbf{F}^{\mathrm{AI}})$ with a Gibbs sampler whose full conditionals are von Mises-Fisher (vMF) distributions on the respective spheres \cite{banerjee2005clustering}. At each iteration,
\begin{align}
\mathbf{h}_X &= \frac{\beta^{\mathrm{human}}}{\sqrt{n_{\mathrm{people}} + n_{\mathrm{fh}}}} J^{\mathrm{human}} \mathbf{F}^{\mathrm{human}} \nonumber \\
&\quad + \frac{\beta^{\mathrm{AI}}}{\sqrt{n_{\mathrm{people}} + n_{\mathrm{fAI}}}} J^{\mathrm{AI}} \mathbf{F}^{\mathrm{AI}}, \nonumber \\
\mathbf{X} &\sim \mathrm{vMF}\!\left( \mathbf{h}_X / \lVert \mathbf{h}_X \rVert, \; \lVert \mathbf{h}_X \rVert \sqrt{n_{\mathrm{people}}} \right),
\label{eq:gibbsX}
\end{align}
and analogously
\begin{align}
\mathbf{h}_{F_h} &= \frac{\beta^{\mathrm{human}}}{\sqrt{n_{\mathrm{people}} + n_{\mathrm{fh}}}} (J^{\mathrm{human}})^{\top} \mathbf{X}, \nonumber \\
\mathbf{F}^{\mathrm{human}} &\sim \mathrm{vMF}\!\left( \mathbf{h}_{F_h} / \lVert \mathbf{h}_{F_h} \rVert, \; \lVert \mathbf{h}_{F_h} \rVert \sqrt{n_{\mathrm{fh}}} \right),
\label{eq:gibbsFh}
\end{align}
\begin{align}
\mathbf{h}_{F_a} &= \frac{\beta^{\mathrm{AI}}}{\sqrt{n_{\mathrm{people}} + n_{\mathrm{fAI}}}} (J^{\mathrm{AI}})^{\top} \mathbf{X}, \nonumber \\
\mathbf{F}^{\mathrm{AI}} &\sim \mathrm{vMF}\!\left( \mathbf{h}_{F_a} / \lVert \mathbf{h}_{F_a} \rVert, \; \lVert \mathbf{h}_{F_a} \rVert \sqrt{n_{\mathrm{fAI}}} \right).
\label{eq:gibbsFa}
\end{align}
After a burn-in period and thinning, we record the overlap
\begin{equation}
O_t := \frac{\langle \mathbf{F}^{\mathrm{human}}, \mathbf{F}^{\mathrm{AI}} \rangle}{\lVert \mathbf{F}^{\mathrm{human}} \rVert \, \lVert \mathbf{F}^{\mathrm{AI}} \rVert} \in [-1, 1],
\label{eq:overlap}
\end{equation}
which captures alignment between the human and AI text representations. For each pair $(\beta^{\mathrm{human}}, \beta^{\mathrm{AI}})$ on the grid $\{0, 0.5, 1.00, 1.25, 1.5, 2.00\}^2$ we estimate the distribution of $O_t$ with Gaussian KDE, detect multimodality with a peak-finding algorithm, and assess normality with the Shapiro-Wilk test \cite{gonzalez2019shapiro} (Fig.~\ref{fig:modelclust}).

\subsection{Statistical analysis}\label{sec:stats}

Survey detection is compared with a binomial process with success probability $0.5$; per-temperature accuracies carry Wilson 95\% confidence intervals (CIs), and deviation from unbiased guessing is quantified by the absolute difference between empirical accuracy and 50\% (Appendix~A). The relation between assigned rating and correct classification (Fig.~\ref{fig:judges}(A)) is estimated by computing, for each rating value 1-10 and each provenance class, the mean and standard error of the correct-classification indicator, then fitting a logistic regression across rating values. For the 2026 wave, group comparisons between Dartmouth and non-Dartmouth respondents use Welch's two-sample $t$ test on per-respondent scores, complemented by the Mann-Whitney $U$ test; each group's departure from chance uses an exact binomial test on the pooled item responses.

\section{Acknowledgments}
L.H. thanks Donny Flynn, Dustin Mixon, and Joe Tien for helpful discussions. All authors thank Dimitrios Giannakis and Kimberly A. Lyford for their help with the paperwork for the Committee for the Protection of Human Subjects (CPHS). L.H. is grateful to Carter Brown, Tony Cicerone, Noah Csank, David Sheng, John Bradley for distributing the surveys. All authors are thankful to the survey respondents and to the Dartmouth students who provided human-written flitzes for the study.

\section{Competing interest}
The authors declare that they have no competing interests.

\section{Funding}
N.C., J.G., S.Z., and Z.Y. are thankful for funding from Undergraduate Research Assistantships at Dartmouth (URAD) with L.H. being their faculty advisor on the grants. L.H., J.G., S.Z., and Z.Y. are thankful for the travel funds to present our work at the Institute for Computational and Experimental Research in Mathematics (ICERM).  

\section{Author contributions}
L.H. designed and supervised the project. J.G., Z.Y., S.Z., and L.H. collected the 2025 data. N.C. and L.H. collected the 2026 data. All authors wrote code and wrote the manuscript. L.H., J.G., S.Z., and N.C. performed analyses and produced figures. L.H. and N.C. revised the manuscript.

\section{Data availability}
The data and code underlying this article are available at \url{https://github.com/lhuynhm/SpinGlasses_LLMs_TuringPhaseTransitions}.

\section{Ethics statement}
All authors completed the Collaborative Institutional Training Initiative (CITI) Human Research course prior to conducting any research with human subjects, and the study was reviewed through the Dartmouth Committee for the Protection of Human Subjects (CPHS). Survey participation was voluntary and anonymous, and the students who contributed human-written flitzes did so with consent.

\section{Declaration on artificial intelligence usage}
J.G., Z.Y., S.Z., and L.H. used ChatGPT-4o to assist with code syntax, while developing the algorithms on their own. N.C. used Claude Opus 5 and Fable 5 to identify relevant references and to assist with writing and coding. The authors developed the research ideas and methodologies on their own.
\newpage
\bibliographystyle{oup-plain}
\bibliography{flitz-bibliography}

\begin{appendices}

\section{Appendix A: Quantification of bias in the 2025 survey data}\label{app:bias}
\setcounter{figure}{0}
\renewcommand{\thefigure}{S\arabic{figure}}

\begin{figure}[!h]%
\centering
\includegraphics[width=\columnwidth]{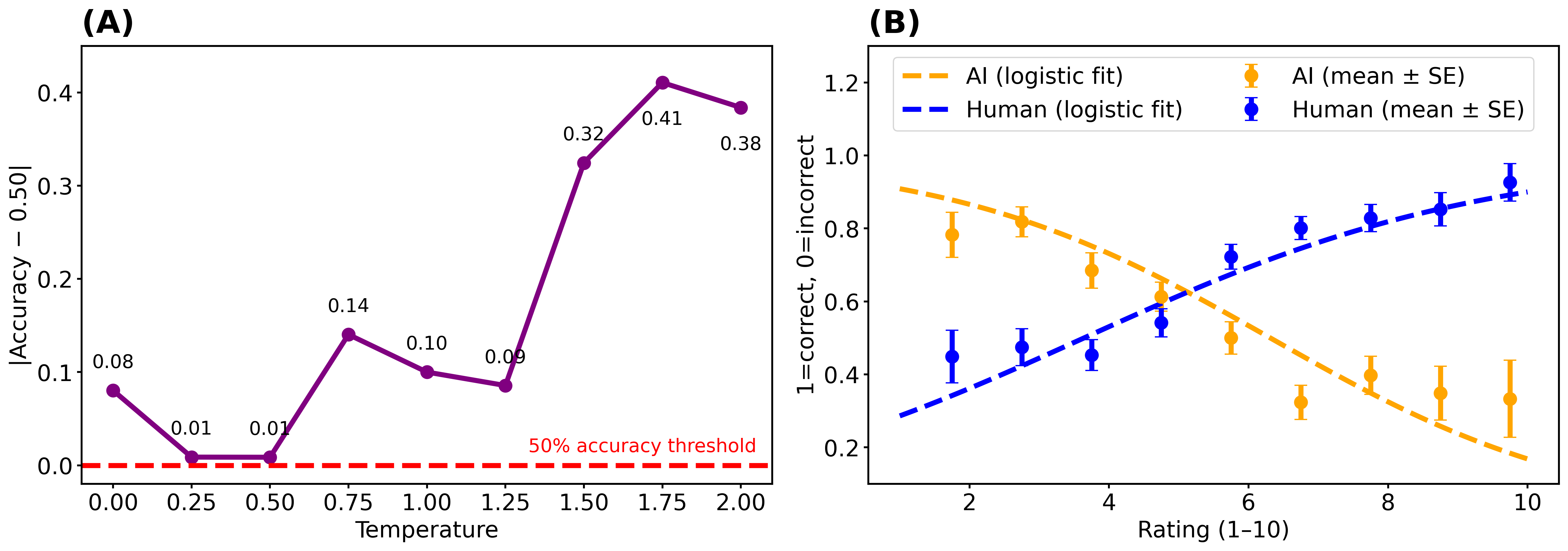}
\caption{Quantification of bias in the 2025 survey responses. (A) Absolute deviation of classification accuracy from the unbiased 50\% baseline at each temperature. (B) Mean correct-classification rate by assigned rating with logistic fits, separated by true provenance (human-written, orange; AI-generated, blue).}\label{fig:bias}
\end{figure}

If respondents held no bias in distinguishing AI-generated from human-written flitzes, their classification behavior would be equivalent to flipping an unbiased coin, and the number of correct guesses at each temperature would be binomial with success probability 50\%. Fig.~\ref{fig:bias}(A) quantifies bias as the absolute deviation of empirical accuracy from 50\% across the nine survey temperatures. Deviation is smallest at $T = 0.25$ and $T = 0.5$, grows with temperature, and jumps most sharply between $T = 1.25$ and $T = 1.5$, peaking at $T = 1.75$, an independent signature of the phase transition located in the Results. Fig.~\ref{fig:bias}(B) disentangles the rationale behind respondents' decisions by regressing correct identification on perceived quality separately for human-written and AI-generated flitzes, yielding the positive and negative logistic trends discussed in the Results: respondents treat quality as evidence of human authorship in both directions.

\section{Appendix B: Pairwise similarity within the human and AI texts}\label{app:heatmaps}

\begin{figure}[!t]%
\centering
\includegraphics[width=\columnwidth]{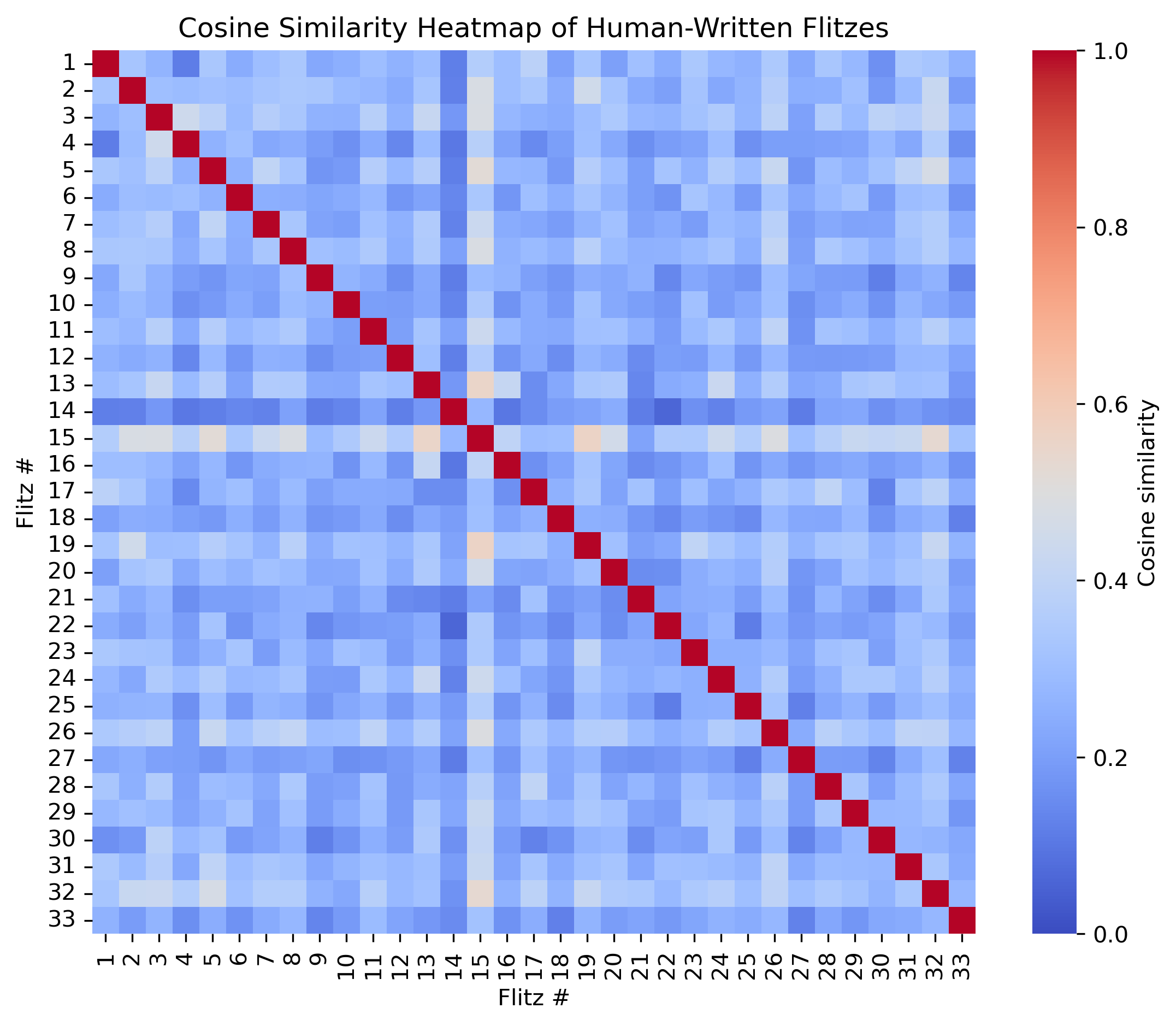}
\caption{Pairwise cosine similarity between the 33 human-written flitzes in the 2025 texts.}\label{fig:hhcos}
\end{figure}

\begin{figure}[!t]%
\centering
\includegraphics[width=\columnwidth]{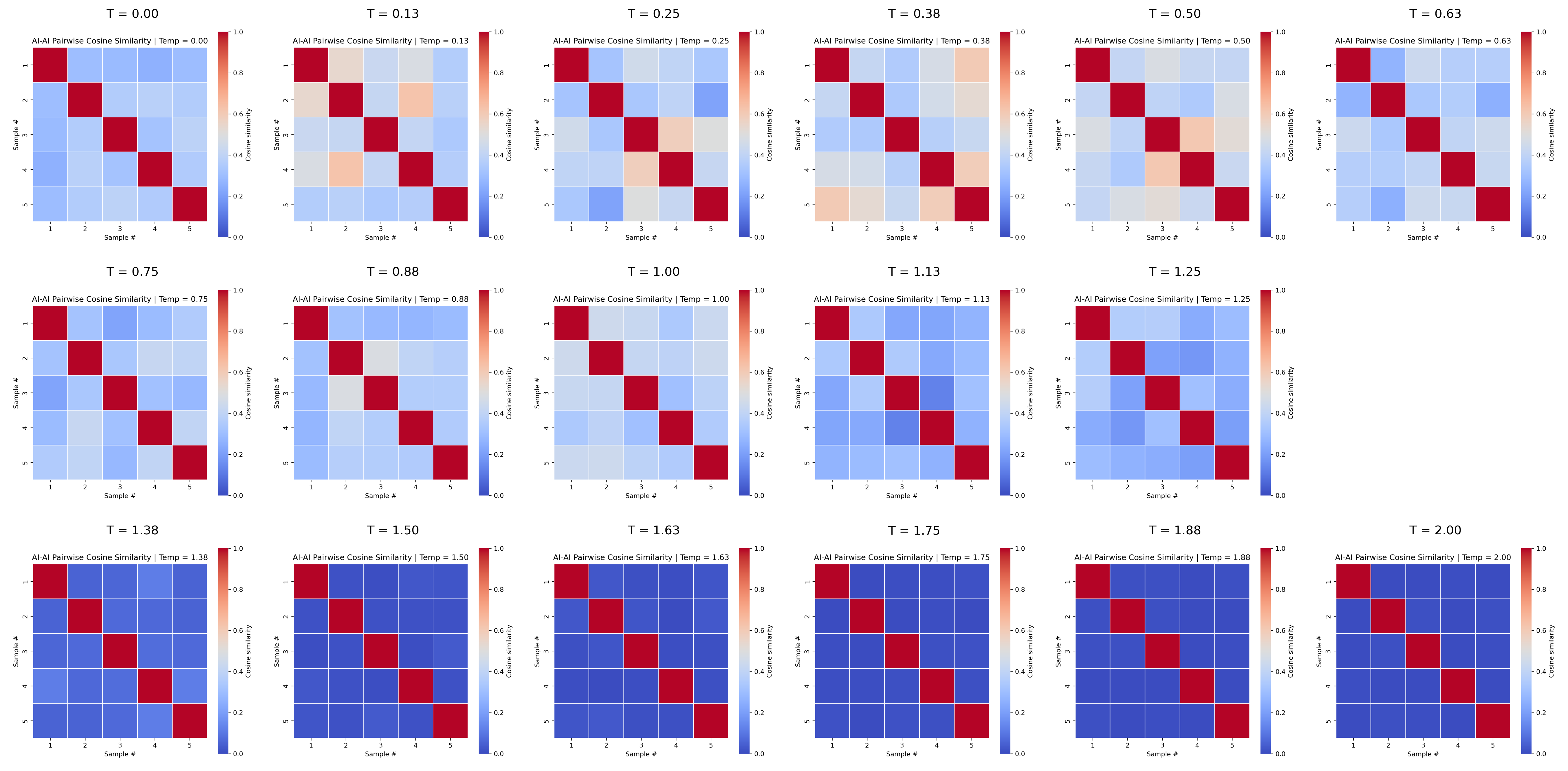}
\caption{Pairwise cosine similarity between randomly selected AI-generated flitzes at each temperature value.}\label{fig:aacos}
\end{figure}

Figure~\ref{fig:hhcos} reports the pairwise TF-IDF cosine similarity between the 33 human-written flitzes. Although a few pairs reach similarity near 0.5 (flitzes 13 and 15, 15 and 19, and 15 and 32), the majority of pairwise scores fall below approximately 0.4, indicating substantial diversity of language across human authors. Figure~\ref{fig:aacos} reports pairwise similarities among randomly selected AI-generated flitzes at each temperature: moderate similarity persists from $T = 0$ through $T = 1.25$, and at $T = 1.38$ the similarity transitions toward zero, matching the phase transition located in the main text.

\end{appendices}

\end{document}